\documentclass[sort&compress]{elsarticle}
\usepackage[top=1in, bottom=1in, left=0.5in, right=0.5in]{geometry}
\usepackage{graphicx}
\usepackage{subfigure}
\usepackage{amssymb}
\usepackage{amsmath}
\usepackage{float}

 \usepackage{lineno}

\journal{Physical Review A}

\begin{document}
\begin{frontmatter}
\bibliographystyle{unsrt}


\title{Practical Attacks on Decoy State Quantum Key Distribution Systems with Detector Efficiency Mismatch }

\author[]{Fei Yangyang}
\author[]{Gao Ming\corref{cor1}}
\author[]{Wang Weilong}
\author[]{Li Chaobo}
\author[]{Ma Zhi}
\cortext[cor1]{Corresponding author: Email: gaoming.zhengzhou@gmail.com}
\address{State Key Laboratory of Mathematical Engineering and Advanced Computing,
Zhengzhou, Henan, 450001, China}
\begin{abstract}\par
To the active-basis-choice decoy state quantum key distribution systems with detector efficiency mismatch, we present a modified attack strategy, which is based on faked states attack, with quantum non-demolition measurement ability to restress the threat of detector efficiency mismatch. Considering that perfect quantum non-demolition measurement ability dosen't exist in real life, we also propose a practical attack strategy using photon number resolving detectors. Theoretical analysis and numerical simulation results show that, without changing the channel, our attack strategies are serious threats to decoy state quantum key distribution systems. The eavesdropper may get some information about the secret key without causing any alarms. Besides, the lower-bound of detector efficiency mismatch to run our modified faked states attack successfully with perfect quantum non-demolition measurement ability is also given out, which provides the producers of quantum key distribution systems with a reference and can be treated as the approximate secure bound of detector efficiency mismatch in decoy state quantum key distribution systems.
\end{abstract}

\end{frontmatter}


\section{Introduction}
\label{}
Quantum key distribution (QKD) ensures the security of secret key exchange between two separated parties, known as Alice and Bob. It is based on the fundamental laws of physics and has been proved to be unconditionally secure \cite{shor2000simple,deutsch1996quantum}. However, the devices of practical QKD systems are not perfect. The eavesdropper, known as Eve, can always take advantage of these loopholes to get information about the secret key. One loophole happens in the weak coherent state source. There may be more than one photon in a pulse, which leads the system vulnerable to an attack named photon number splitting (PNS) \cite{huttner1995quantum}. Although it is nearly impossible to perform perfect PNS attack with prior technology, it is still a big potential threat to the practical QKD systems. In 2011, a simplified PNS attack was implemented successfully \cite{liu2011proof}. To resist PNS attack, the idea of decoy state method was proposed \cite{hwang2003quantum,wang2005beating,lo2005decoy,ma2005practical}, which is widely used in practical QKD systems. \par

Another loophole named detector efficiency mismatch (DEM) widely exists in practical QKD systems. There are always at least two separate gated single photon avalanche detectors for ``0'' and ``1'' values on Bob's side. In the ideal model, the two detectors' efficiency curves are assumed to be perfectly matched. However, it is not like that in real life. There is a probability of approximately 4\% that large DEM occurs in practical QKD systems \cite{zhao2008quantum}. Besides, Eve can also induce a large temporal DEM by interfering the calibration of the detectors \cite{jain2011device}.\par
Faked states attack (FSA) \cite{makarov2005faked} is an intercept-resend attack, which works due to the existence of large DEM. Eve randomly chooses her measurement basis, then prepares the opposite bit value in the opposite basis according to her measurement results and resends the faked states to Bob at different time, denoted as ${t_0}$ and ${t_1}$. Bob's detectors barely click when his measurement basis choice is different from Eve's. In reality, to maintain Bob's overall detection probability the same as that before mounting this attack, Eve uses a weak coherent state source and increases the brightness of her faked states \cite{jain2011device,makarov2006effects}. \par
Time-shift attack also exploits DEM \cite{qi2005timeshift,zhao2008quantum}. In this attack, Eve shifts the arrival time of each of Alice's pulse forward or backward as she wishes. By this method, Eve can determine the detection results with large probability and introduce no additional error. However, comparing with FSA, one drawback of time-shift attack is that it can only compensate the decrease of Bob's detection probability by changing the transmission of the channel, which means that Eve may be discovered when the transmission distance is too short. Besides, changing the transmission of the channel is not as convenient as changing the brightness of the faked states in FSA.\par
Insufficient models of single photon detectors are also serious threats to the security of practical QKD systems, which can be attacked by blinding the detectors with bright illumination \cite{lydersen2010hacking,lydersen2010thermal,sauge2011controlling,makarov2009controlling,lydersen2011controlling}. When the detectors are blinded, they only respond to the bright pulses. In a modified intercept-resend attack with strong resent pulses, the eavesdropper can control the responses of the detectors to get a full copy of the secret key. This kind of attacks would be a serious concern for practical QKD systems. Fortunately, protections against bright illumination attacks on gated avalanche photodiodes by correctly operating them were proposed in 2011 \cite{yuan2011resilience}.\par
In Ref. \cite{makarov2006effects}, it gave out an idea to mount FSA on weak coherent state source with quantum non-demolition (QND) measurement ability. However, perfect QND measurement ability doesn't exist in reality. To make FSA on decoy state QKD more practical, we can use photon number resolving detectors (PNRDs) instead. \par
PNRDs are widely used in linear optical quantum computing \cite{knill2001scheme} and quantum information processing \cite{hwang2003quantum,ekert2000quantum}. There are several types of mechanisms used to construct PNRDs \cite{peacock1998recent,shields2000detection,divochiy2008superconducting,rosenberg2005noise,jiang2007photon}. Each mechanism has its own advantages and disadvantages \cite{eisaman2011invited}. Without 100\% single-photon detection efficiency, PNRDs can not tell the exact photon number of each incident pulse and the measured photon number is just a lower estimate. The single-photon detection efficiency should be as large as possible in order to maximize the probability of detecting all the photons in the incident pulse. \par

 FSA on single photon QKD was studied in Ref. \cite{makarov2006effects}, but quantitative analysis on decoy state QKD has not been done yet. Security of QKD systems with DEM was analyzed in \cite{fung2009security,DEM2010,maroy2010security}. The results showed that DEM must be bounded to ensure the security of QKD systems. In Ref. \cite{DEM2010}, the secure key generation rate formula that took DEM into account was provided and the secure bound of DEM in single photon QKD was also given out. However, the secure bound of DEM in decoy state QKD has not been presented because of its complexity. What's more, the secure key generation rate formula presented in Ref. \cite{DEM2010} is not practical in reality because it will decrease the key generation rate.\par
 In this paper, we attack decoy state QKD systems with DEM to restress the threat of DEM. First, we present a modified strategy, which is based on FSA, with the ability to do QND measurement. We also provide the lower-bound of DEM to run FSA successfully on decoy state QKD systems, which can be treated as the approximate secure bound of DEM in decoy state QKD. Since perfect QND measurement ability doesn't exist in real life, we find that it's also possible to perform our attack strategy with PNRDs. The attack strategy is exemplified using weak + vacuum decoy state BB84 \cite{bennett1984quantum} QKD here.\par

Measurement-device-independent QKD protocol was proposed to defense all side-channel attacks on the loopholes of practical detectors in 2012 \cite{lo2012measurement}. However, this protocol is difficult to realize in real world and its key generation rate (about 10bps at 100km \cite{tang2014measurement}) is much lower than the traditional decoy state BB84 protocol (about 10kbps at 100km \cite{dixon2008gigahertz}), which limits its application in practical QKD systems. So our work to attack on decoy state QKD systems is still meaningful. \par
	This paper is organized as followed. In Sec. 2 we grant Eve a future technology named QND measurement ability to mount FSA on decoy state QKD. The attack strategy is described and numerical simulation is done. We also present the lower-bound of DEM to successfully attack with perfect QND measurement ability, which gives the producers of practical QKD systems a reference. In Sec. 3 we consider a more practical situation that Eve mounts FSA with PNRDs on decoy state QKD. The security of decoy state QKD under this attack is analyzed in Sec. 4. Finally, discussion and conclusion are made in Sec. 5.
\section{Attack with Perfect QND Measurement Ability on Decoy State QKD Systems}
In this section, we will give out an attack strategy based on FSA on decoy state QKD with the assumption that Eve has the ability to do perfect QND measurement. The attack strategy is described, then results of numerical simulation are given out.
\subsection{Attack strategy}
If Eve wants to perform attack on decoy state QKD systems successfully, one possible way is to keep the key generation rate $R$ and the overall detection probability  ${Q_\mu}$ close to the data before mounting the attack, so QBER is naturally lower than the threshold at the same time. In this way, Eve can get information about the secret key while she is hidden. \par
Fig. 1 shows the simple diagram of the attack strategy. With perfect QND measurement ability, Eve can get the photon number information of every incident signal. According to the measured photon number, Eve controls the optical switch to mount FSA on those signals that contain only one photon.
 \begin{figure}[H]
   \centering
   \includegraphics[width=0.6\textwidth]{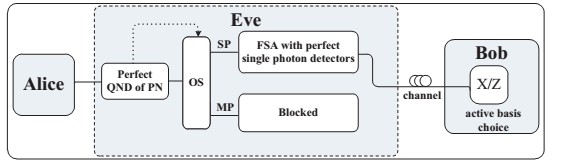}
   \caption{Simple diagram of our attack strategy that Eve has QND measurement ability. PN, photon number; OS, optical switch; SP, single-photon signal; MP, multi-photon signal; X/Z, active basis choice.}\label{fig:digit}
 \end{figure}
Usually, those signals that contain more than one photon can be passed undisturbedly to Bob or eavesdropped on using PNS attack or just blocked. In Ref \cite{makarov2006effects}, it proposed the idea to mount FSA on weak coherent state source by letting all multi-photon signals be passed undisturbedly. But we find that it's the best way for Eve to block all the multi-photon signals. The benefits of this strategy are that, first of all, the secure key of QKD with weak coherent state source all comes from signals that contain no more than one photon. Blocking all the multi-photon signals will increase the lower-bound of the gain of single-photon signals, ${Q_1}^L$. We can get this conclusion by expanding Eq. 35 in ref. \cite{ma2005practical}. The expansion is as follow:
\[{Q_1}^L = \frac{{{\mu ^2}{e^{ - \mu }}}}{{\mu \nu  - {\nu ^2}}}\sum\limits_{i = 1}^\infty  {{Y_i}\frac{{{\nu ^2}({\nu ^{i - 2}} - {\mu ^{i - 2}})}}{{i!}}} \]
where ${\mu}$ is the average photon number of signal state, ${\nu}$ is the average photon number of decoy state, and ${\nu}$${\textless}$${\mu}$. ${Y_i}$  is the yield of an $i$-photon pulse. It is easy to get ${Q_1}^L$ diminishing with $i$, because ${Y_i}\frac{{{\nu ^2}({\nu ^{i - 2}} - {\mu ^{i - 2}})}}{{i!}}$ is non-positive when $i$${\geq}$2. Blocking all the multi-photon signals makes $i$=1 or 0, thus increases ${Q_1}^L$. A higher ${Q_1}^L$ leads to a higher secure key generation rate which is good for Eve. Second, the idea of letting those multi-photon signals be passed undisturbedly will make a small part of the key unavailable, but our strategy will not, which is the same as the idea of eavesdropping the multi-photon signals using PNS attack.\par
In order to maintain the overall detection probability of Bob's detectors, Eve uses weak coherent state source to generate faked states. The average photon number of the faked states sent by Eve at ${t_0}$ (${t_1}$) is ${\mu_0}$(${\mu_1}$). According to Ref. \cite{jain2011device}, the detection probability of faked states on Bob's detector ``0'', ${p_0({\mu_0},{\mu_1})}$, and ``1'', ${p_1({\mu_0},{\mu_1})}$, are:
\[{p_0}({\mu _0},{\mu _1}) = 0.75 + 0.25d - 0.25(1 - d)({e^{ - 0.5{\mu _0}{\eta _{00}}}} + {e^{ - 0.5{\mu _1}{\eta _{01}}}} + {e^{ - {\mu _1}{\eta _{01}}}}),(1)\]
\[{p_1}({\mu _0},{\mu _1}) = 0.75 + 0.25d - 0.25(1 - d)({e^{ - 0.5{\mu _0}{\eta _{10}}}} + {e^{ - 0.5{\mu _1}{\eta _{11}}}} + {e^{ - {\mu _1}{\eta _{10}}}}),(2)\]
where ${\eta_{mn}}$, $m$,$n$${\in}$${\{0,1\}}$, represents the equivalent overall transmission and detection efficiency between Alice and Bob's detector $m$ at time ${t_n}$. $d$ is the dark count probability of Bob's detectors.\\
The total detection probability of faked states on Bob's detectors ${p_{arrive}({\mu_0},{\mu_1})}$ is\\
\[\begin{array}{c}
{p_{arrive}}({\mu _0},{\mu _1}) = 1 - 0.25(1 - d)({e^{ - {\mu _1}{\eta _{01}}}} + {e^{ - {\mu _0}{\eta _{10}}}}) + 0.25d(1 - d)\\
 \times ({e^{ - {\mu _1}{\eta _{01}}}} + {e^{ - {\mu _0}{\eta _{10}}}}) - 0.25{(1 - d)^2}({e^{ - 0.5{\mu _0}{\eta _{00}} - 0.5{\mu _0}{\eta _{10}}}} + {e^{ - 0.5{\mu _1}{\eta _{01}} - 0.5{\mu _1}{\eta _{11}}}}).
\end{array}\]
And the error rate of faked states ${p_{error}({\mu_0},{\mu_1})}$ is\\
\[\begin{array}{c}
{p_{error}}({\mu _0},{\mu _1}) = 0.125(1 - d)({e^{ - 0.5{\mu _0}{\eta _{00}}}} + {e^{ - 0.5{\mu _1}{\eta _{11}}}} - {e^{ - 0.5{\mu _0}{\eta _{10}}}} - {e^{ - 0.5{\mu _1}{\eta _{01}}}} - {e^{ - {\mu _1}{\eta _{01}}}} - {e^{ - {\mu _0}{\eta _{10}}}})\\
 - 0.125{(1 - d)^2}({e^{ - 0.5{\mu _0}{\eta _{00}} - 0.5{\mu _0}{\eta _{10}}}} + {e^{ - 0.5{\mu _1}{\eta _{01}} - 0.5{\mu _1}{\eta _{11}}}}) + 0.125d(1 - d)({e^{ - {\mu _1}{\eta _{01}}}} + {e^{ - {\mu _0}{\eta _{10}}}}){\rm{ + }}0.5.
\end{array}\]
In our attack strategy with QND measurement ability, the overall detection probability of signal state on Bob's detectors includes the detection probability of faked states resent by Eve and the dark count probability, that is
\[{Q_\mu } = {p_{arrive}}({\mu _0},{\mu _1})\mu {e^{ - \mu  }} + (1 - \mu {e^{ - \mu}})d.\]
Similarly the detection rate of the decoy state with an average photon number of ${\nu}$ is
\[{Q_\upsilon } = {p_{arrive}}({\mu _0},{\mu _1})\upsilon  {e^{ - \upsilon  }} + (1 - \upsilon  {e^{ - \upsilon  }})d.\]
The error rate contains the error probability of faked states and the error probability from dark count, so we have
\[{E_\mu }{Q_\mu } = {p_{error}}({\mu _0},{\mu _1})\mu  {e^{ - \mu  }} + \frac{1}{2}(1 - \mu  {e^{ - \mu  }})d.\]
And the overall QBER after attack is given by
\[{E_\mu } = \frac{{{E_\mu }{Q_\mu }}}{{{Q_\mu }}} .\]
Eve is able to control three parameters. They are ${\mu_0}$, ${\mu_1}$ and the maximum value of DEM denoted as ${k}$, where ${k}$=${\eta_{00}}$/${\eta_{10}}$=${\eta_{11}}$/${\eta_{01}}$ and ${\eta_{00}}$=${\eta_{11}}$. In order to keep the detection probability of Bob's ``0'' and ``1'' detectors equivalent, according to Eqs. (1), (2), we can assume that  ${\mu _0}$=${\mu _1}$=${\mu '}$. Alice and Bob use the following formulas
\[{Y_1}^L = \frac{\mu }{{\mu \nu  - {\nu ^2}}}({Q_\nu }{e^\nu } - {Q_\mu }{e^\mu }\frac{{{\nu ^2}}}{{{\mu ^2}}} - \frac{{{\mu ^2} - {\nu ^2}}}{{{\mu ^2}}}d),\]
\[{Q_1}^L = \mu {e^{ - \mu }}{Y_1}^L,\]
\[{e_1}^U = \frac{{{E_v}{Q_v}{e^v} - \frac{1}{2}d}}{{{Y_1}^L\nu }},\]
to estimate the lower-bound of the gain of single-photon signals ${Q_1}^L$ and the upper-bound of the error rate of single-photon signals ${e_1}^U$. According to the idea of GLLP, they can get the lower-bound of the key generation rate $R$, which is given by
\[R \ge q\{  - {Q_\mu }f({E_\mu }){H_2}({E_\mu }) + Q_1^{L}[1 - {H_2}(e_1^{U})]\}, \]
where ${q=\frac{1}{2}}$, $f(x)$ is the bidirectional error correction efficiency, ${H_2}(x) =  - xlo{g_2}x - (1 - x)lo{g_2}(1 - x)$ is the binary Shannon information function.
\subsection{Numerical simulation}
The numerical simulations in this paper use some GYS \cite{gobby2004quantum} experiment parameters, including the loss coefficient in the quantum channel ${\alpha=0.21 }$dB/km; the dark count probability $d$=${1.7\times10^{-6}}$; the transmittance in Bob's side ${\eta _{Bob}}$ =0.045; the average photon number of the signal state ${\mu=0.48}$; the average number of the decoy state ${\nu=0.05}$; the bidirectional error correction efficiency is 1.22. We also assume that $k$${\leq}1000 $, ${\eta _{01}}$=${t _{AB}}$${\eta _{Bob}}$${\times}$${10^{-4}}$(these two assumptions are reasonable and they can be achieved in reality \cite{jain2011device}), where ${t _{AB}}$ is the channel transmittance. ${e _{detector}}$=0 is also assumed to simplify our calculation processing, which means no photon hits the erroneous detector.\par
\begin{figure}[H]
  \centering
  \includegraphics[width=0.6\textwidth]{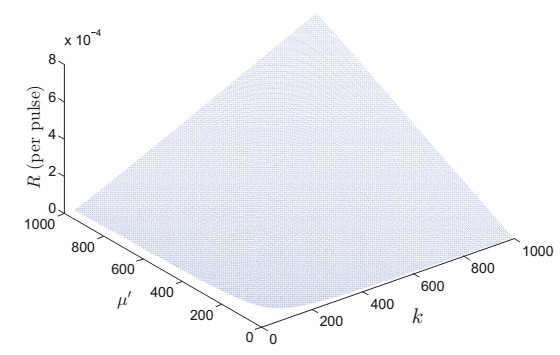}
  \caption{The relationship of $R$, $k$ and ${\mu '}$ when Eve has perfect QND measurement ability.}\label{fig:digit}
\end{figure}\par
For a fixed transmission distance of 100km, when Eve has perfect QND measurement ability, Fig. 2 shows the relationship of $R$, ${k}$ and ${\mu '}$ when $R$ is positive and it is ignored when ${R}$${\textless}0$ .
We can figure out that $R$ rises along with ${k}$ and ${\mu '}$, and there are many combinations of ${k}$ and ${\mu '}$ for the same $R$. Besides, when $k$ is small enough, $R$ is negative no matter how large ${\mu'}$ is.

 So it is easy to find a tuple [${k}$, ${\mu '}$] that makes our attack successful with perfect QND measurement ability, such as ${k}$=310, ${\mu '}$=300. Fig. 3 gives out the comparisons of $R$ and ${Q_{\mu}}$.
 \begin{figure}[H]
\centering
\subfigure[]{
\label{Fig.sub.1}
\includegraphics[width=0.5\textwidth]{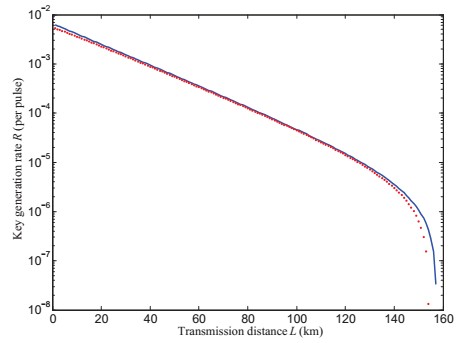}}
\subfigure[]{
\label{Fig.sub.2}
\includegraphics[width=0.5\textwidth]{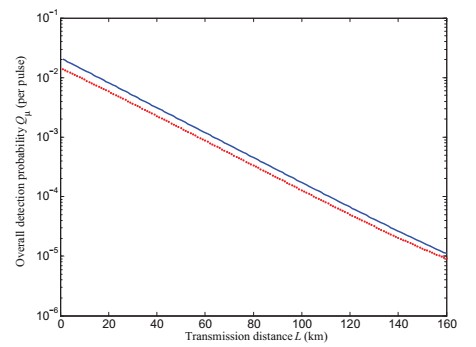}}
\caption{(color online). (a)The solid line shows the key generation rate without attack. The dotted line shows the key generation rate under our attack with perfect QND measurement ability. Here assuming that ${e _{detector}}$=0, so the secure transmission distance extends from 140km to 160km. (b)The solid line shows the detection probability of signal state without attack. The dotted line shows the detection probability of signal state under our attack with perfect QND measurement ability. We have to note that legitimate users do not take DEM into account when calculating the key generation rate and the detection probability without attack.}
\label{Fig.lable}
\end{figure}
\par
From Fig. 3, we can see that when Eve has perfect QND measurement ability, $R$ and ${Q_{\mu}}$ under our attack are both very close to the normal value. So Eve stays undetected. We have to notice that the numerical simulation above uses ${k}$=310, not the maximum value 1000. As we enlarge the value of ${k}$, the attack effects will be better.\par
Although the lager ${k}$ is, the better attack effects Eve will get, we are still concerned about the minimum value of ${k}$, which is set as ${k_{min}}$, to make $R$ positive in ideal situation that Eve has perfect QND measurement ability. Fig. 4 shows the relationship between ${k_{min}}$ and the transmission distance $L$. ${k_{min}}$ rises with $L$, and when ${k_{min}}$ is larger than 35, FSA on decoy state QKD is possible in ideal situation. This provides a reference for the producers of QKD systems. If they can improve the calibration process to guarantee the DEM below 35, FSA will no longer be a threat to decoy state QKD. In other words, the result of 35 can also be treated as the approximate secure bound of DEM in decoy state QKD systems.
\begin{figure}[H]
  \centering
  \includegraphics[width=0.6\textwidth]{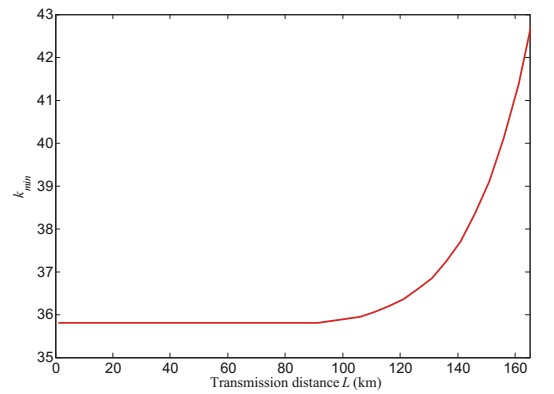}
  \caption{The relationship between ${k_{min}}$ and $L$}\label{fig:digit}
\end{figure}
\section{Practical Attack with PNRDs on Decoy State QKD Systems}
In Sec. 2, we consider the situation that Eve has perfect QND measurement ability, which doesn't exist in the real world. In this section, we will discuss a more realistic situation that Eve only has PNRDs. We present our attack strategy first, and show numerical simulation results later.
\subsection{Attack strategy}
\begin{figure}[H]
  \centering
  \includegraphics[width=0.6\textwidth]{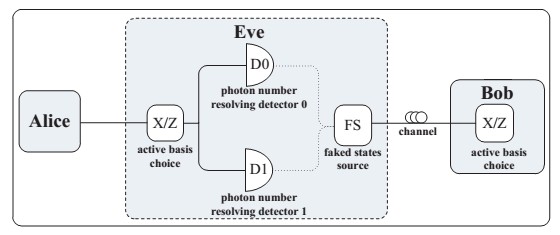}
  \caption{Simple diagram of our attack strategy with PNRDs. Eve and Bob use active basis choice here. X/Z, active basis choice; D0, photon number resolving detector 0; D1,
  photon number resolving detector 1; FS, faked states source.}\label{fig:digit}
\end{figure}
Fig. 5 shows the simple diagram of our practical attack strategy with PNRDs. Eve attacks the system with active basis choice at a place close to Alice. She uses two PNRDs for ``0'', ``1'' bit value, and gets the ``photon number'' of every incident pulses by calculating the summation of the detection results of two detectors, which can not be achieved by using single photon detectors. In our attack strategy, Eve mounts FSA only when her measurement results are single-photon signals, and the corresponding probability of signal state is \[{{p_{single}}={\sum\limits_{{\rm{i}} = 1}^\infty  {p(i)\left( {\begin{array}{*{10}{c}}
i\\
1
\end{array}} \right)\eta (1 - \eta ) ^{i - 1}}}=\mu}{\eta}{e^{-\mu\eta}},\] where ${\eta}$ is the single-photon detection efficiency of Eve's PNRDs, $p(i) = \frac{{{\mu ^i}}}{{i!}}{e^{ - \mu }}$ is the probability of $i$-photon signal. The dark count probability of Eve's PNRDs is ignored because it's much smaller than ${p_{single}}$. There are several benefits of this strategy. First, this strategy may partially distinguish
the multi-photon signals and block them. Second, Eve can get most of the information about the secret key, and this strategy is easy to perform. At last, ``double-click'' means that Eve's basis choice is different from Alice's, and it should be discarded to benefit Eve. Mounting FSA only when Eve's measurement results are single-photon signals eliminates the influence of ``double-click''.\par
Similar with the attack strategy in Sec. 2, we can get the overall detection probability of signal state on Bob's detectors:
\[{Q_\mu } = {p_{arrive}}({\mu _0},{\mu _1})\mu \eta {e^{ - \mu \eta }} + (1 - \mu \eta {e^{ - \mu \eta }})d.\]
The detection rate of the decoy state with an average photon number of ${\nu}$ is
\[{Q_\upsilon } = {p_{arrive}}({\mu _0},{\mu _1})\upsilon \eta {e^{ - \upsilon \eta }} + (1 - \upsilon \eta {e^{ - \upsilon \eta }})d.\]
The error rate is
\[{E_\mu }{Q_\mu } = {p_{error}}({\mu _0},{\mu _1})\mu \eta {e^{ - \mu \eta }} + \frac{1}{2}(1 - \mu \eta {e^{ - \mu \eta }})d.\]
And the overall QBER after attack is given by
\[{E_\mu } = \frac{{{E_\mu }{Q_\mu }}}{{{Q_\mu }}} .\]
\subsection{Numerical simulation}
 Here we take ${k}$=1000, ${\mu '}$=900 and ${\eta}$=0.1 \cite{kardynal2008avalanche}. Fig. 6 gives out the comparisons of $R$ and ${Q_{\mu}}$ when attacking with PNRDs.
  \begin{figure}[H]
\centering
\subfigure[]{
\label{Fig.sub.1}
\includegraphics[width=0.5\textwidth]{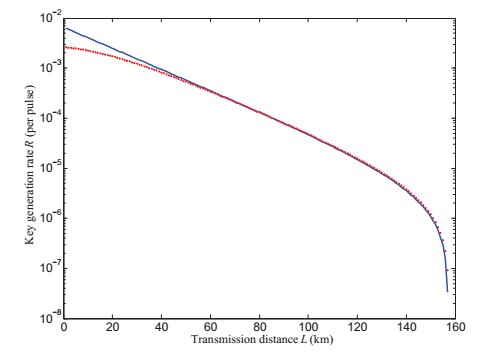}}
\subfigure[]{
\label{Fig.sub.2}
\includegraphics[width=0.5\textwidth]{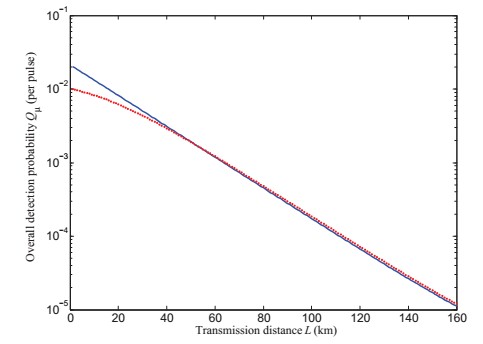}}
\caption{(color online). (a)The solid line shows the key generation rate without attack. The dotted line shows the key generation rate under our attack with PNRDs that ${\eta}$=0.1. (b)The solid line shows the detection probability without attack. The dotted line shows the detection probability under our attack with PNRDs that ${\eta}$=0.1.}
\label{Fig.lable}
\end{figure}
 As shown in Fig. 6, when ${L}$${\textgreater}30$ km, $R$ and ${Q_{\mu}}$ under our attack are both very close to the normal value. So Eve remains hidden. The deviation from theoretical values is large only when the transmission distance is shorter than 30km. This is because the channel transmittance is large when the distance is short, which makes the expectation detection probability large. What's more, large average photon number of faked states causes large overall error rate, and makes $R$ lower than legitimate users' expectation. \par
 The measurements in the attack strategy with perfect QND ability can be separated into two parts. The first one is the photon number measurement, and the other is the measurement when mounting FSA. However, our strategy with practical PNRDs can put these two parts together which measures the photon number and does the measurement in FSA at the same time. It is also the reason why our strategy is interesting. When ${\eta}$=1, Eve is able to mount FSA on all the single-photon signals and block all the multi-photon signals which is exactly the same with that using perfect QND measurement ability.\par
\section{The Security Analysis}
Eve's attack with QND measurement ability is always better than that with PNRDs. When Eve attacks with PNRDs whose ${\eta}$ is not 1, she can only distinguish part of the single-photon signals.  To maintain ${Q_\mu}$ the same as Bob's expectation, the average photon number of faked states should be larger than that attacking with perfect QND measurement ability. However, larger average photon number of the faked states leads to larger error rate, which decreases the key generation rate. \par
Here we analyze the security of the strategy that Eve attacks with PNRDs as an example. In our attack strategy, Eve might steal some information about the secret key without being detected. However, she can not get all of the key. When Eve's measurement basis is different from Alice's and Bob's, legitimate users may share some absolutely secure key while Eve doesn't know, which is represented by ${R_{absolute}}$. To explain ${R_{absolute}}$ in detail, we give out an example that Alice prepares bit 0 in the Z basis and Bob measures in the Z basis, while Eve measures in the X basis. When Eve's measurement result is bit 1, according to the FSA theory, the faked state resent by Eve is bit 0 in the Z basis and Bob will get bit 0 as a result. Although the probability of this kind of cases is very small, Alice and Bob will share some absolutely secure key. Using the similar idea in Ref. \cite{jain2011device} we can get Table 1, from which ${R_{absolute}}$ can be calculated.
\begin{table}[H]
\caption{Given that Alice prepares bit 0 and 1 in the Z basis and that Bob
measures in the Z basis, Eve measures in X basis to mount FSA. The first
column contains Alice's bit value.  The second column shows Eve's measurement result. The third
column shows the parameters of the faked state resent by Eve: basis, bit, mean photon number, timing.  The fourth column shows Bob's measurement result; 0 $\cap$ 1 denotes a double click. The last column shows the corresponding detection probabilities. }
\centering  
\scriptsize
\begin{tabular}{ccccc}  
\hline
\hline
Z$\rightarrow$ & {$\rightarrow$Eve} & {Eve$\rightarrow$} & {Bob's result} & {Detection probability}\\ \hline  
0 or 1& 0 & {Z,1,${\mu_0}$,${t_0}$} & 0 & ${r_0=0}$\\
&&&1&${r_1=1-exp[-\mu_0\eta_1(t_0)]}$ \\
&&&${0 \cap 1}$&${r_0r_1}$ \\
&&&Loss&${1-(r_0+r_1-r_0r_1)}$ \\
0 or 1 & 1 & {Z,0,${\mu_1}$,${t_1}$} & {0} & ${s_0=1-exp[-\mu_1\eta_0(t_1)]}$   \\   
&&&1&${s_1=0}$ \\
&&&${0 \cap 1}$&${s_0s_1}$ \\
&&&Loss&${1-(s_0+s_1-s_0s_1)}$ \\
\hline
\hline
\end{tabular}
\end{table}
There are two situations that will induce ${R_{absolute}}$. One is that Alice prepares in the Z basis, Bob measures in the Z basis and Eve measures in the X basis. The other is Alice prepares in X basis, Bob measures in the X basis and Eve measures in the Z basis. The probability that the two situations above occur is ${\frac{1}{4}}$. From Table 1, in the first situation, the absolutely secure key rate is ${\frac{1}{2}(r_1+s_0)}$. Similarly, the absolutely secure key rate in the second situation is the same. The probability that we mount FSA is ${\mu \eta {e^{ - \mu \eta }}}$. So we can get the overall absolutely secure key between Alice and Bob is
\[{R_{absolute}} = \mu \eta {e^{ - \mu \eta }} \times \frac{1}{4} \times \frac{1}{2}({r_1} + {s_0}) = \frac{1}{8}\mu \eta {e^{ - \mu \eta }}({r_1} + {s_0}).\]
 According to the decoy state theory, if the key generation rate ${R}$ is larger than ${R_{absolute}}$, it means that Eve can always get some information about the key. Here we take ${\eta}$=0.1 to do numerical simulation.

The result is shown in Fig. 7. As we can see, ${R_{absolute}}$ is always smaller than ${R}$, which means Eve can always get information about the key while she is hidden. So the overall security of decoy state QKD is broken.
\begin{figure}[H]
  \centering
  \includegraphics[width=0.6\textwidth]{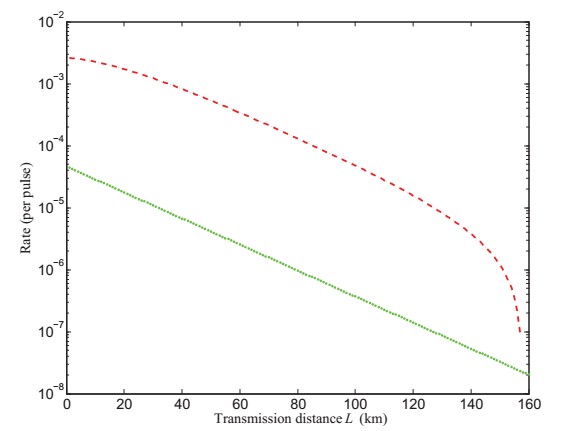}
  \caption{The dashed line represents the key generation rate ${R}$ and the dotted line shows the relationship between ${R_{absolute}}$ and ${L}$.}\label{fig:digit}
\end{figure}
\section{Discussion and Conclusion}
In this paper, to the decoy state QKD systems with active basis choice, we give out a modified attack strategy with perfect QND measurement ability to restress the threat of DEM. We also propose a more practical attack strategy using PNRDs as perfect QND measurement ability dosen't exist in the real world. In our attack strategy, Eve blocks all the multi-photon signals and only mounts FSA when her measurement results are single-photon signals. We find that Eve can maintain the key generation rate $R$ and the detection probability ${Q_{\mu}}$ close to the data without being attacked, and keep the QBER low at the same time. So Eve can remain undetected. The security analysis shows that the eavesdropper could always get information about the key without being detected, so the overall security of practical decoy state QKD systems is broken. We also present the lower-bound of DEM (about 35) to mount FSA successfully with perfect QND measurement ability on decoy state QKD systems, which can be treated as the approximate secure bound of DEM.\par
In conclusion, by mounting our practical attack strategy on the active-basis-choice decoy state QKD systems with no corresponding protections, the eavesdropper is able to get information about the secret key while she is hidden.\\
\section*{Acknowledgments}
This work was supported by the National High Technology Research and Development Program of China
(2011AA010803), the National Natural Science Foundation of China (61472446, U1204602) and the Open
Project Program of the State Key Laboratory of
Mathematical Engineering and Advanced Computing
(2013A14).

\begin{center}
\underline{\hspace{12cm}}
\end{center}
\section*{References}
\bibliographystyle{elsarticle-harv}
\bibliography{bibliography}

\end{document}